\documentclass[10pt]{iopart}

\usepackage{graphicx}
\expandafter\let\csname equation*\endcsname\relax
\expandafter\let\csname endequation*\endcsname\relax

\usepackage{amssymb,amsmath,subfigure,booktabs,bm,multirow,pifont}
\usepackage[ruled]{algorithm2e}
\usepackage[colorlinks,citecolor=green]{hyperref}

\begin{document}

\title[]{User-wise Perturbations for User Identity Protection in EEG-Based BCIs }

\author{Xiaoqing~Chen, Siyang~Li, Yunlu~Tu, Ziwei~Wang, Dongrui~Wu*}

\address{Huazhong University of Science and Technology, Wuhan, China}
\ead{drwu@hust.edu.cn}
\vspace{10pt}
%\begin{indented}
%%\item[]August 2017 (minor update March 2024)
%\end{indented}

\begin{abstract}
\emph{Objective}: An electroencephalogram (EEG)-based brain-computer interface (BCI) is a direct communication pathway between the human brain and a computer. Most research so far studied more accurate BCIs, but much less attention has been paid to the ethics of BCIs. Aside from task-specific information, EEG signals also contain rich private information, e.g., user identity, emotion, disorders, etc., which should be protected. \emph{Approach}: We show for the first time that adding user-wise perturbations can make identity information in EEG unlearnable. We propose four types of user-wise privacy-preserving perturbations, i.e., random noise, synthetic noise, error minimization noise, and error maximization noise. After adding the proposed perturbations to EEG training data, the user identity information in the data becomes unlearnable, while the BCI task information remains unaffected.  \emph{Main results}: Experiments on six EEG datasets using three neural network classifiers and various traditional machine learning models demonstrated the robustness and practicability of the proposed perturbations. \emph{Significance}: Our research shows the feasibility of hiding user identity information in EEG data without impacting the primary BCI task information.
\end{abstract}

\section{Introduction}

A brain-computer interface (BCI) \cite{Wolpaw2002} enables direct communication between the brain and an external device, such as a computer, wheelchair, or robot. It can be used in neurological rehabilitation \cite{Daly2008}, awareness evaluation \cite{Li2015}, active tactile exploration \cite{ODoherty2011}, robotic device control \cite{Hochberg2012}, speech synthesis \cite{Anumanchipalli2019}, etc. The electroencephalogram (EEG), a cost-effective and convenient brain signal measure from the scalp, is the primary input for non-invasive BCIs. There are different paradigms/tasks in EEG-based BCIs, such as motor imagery (MI) \cite{Pfurtscheller2001}, event-related potentials (ERP) \cite{Picton2000}, steady-state visual evoked potential (SSVEP) \cite{Norcia2015}, affective BCI \cite{Wu2023}, etc.

In recent years, multiple laws worldwide have been enacted to address increasing privacy concerns of physiological data, such as the General Data Protection Regulation of the European Union and the Personal Information Protection Law of China \cite{Zhang2022}. They require that users have the right to hide private information in their data. However, EEG data inherently contains sensitive private information irrespective of the specific tasks \cite{Xia2023, Zhang2021a}. Martinovic \emph{et al.} \cite{Martinovic2012} demonstrated that diverse private information, such as credit cards, personal identification number, known individuals, and residential addresses, could be deduced from EEG signals. Furthermore, Icena \emph{et al.} \cite{Ienca2018} highlighted privacy challenges and ethical concerns associated with direct-to-consumer neurotechnologies. Landau \emph{et al.} \cite{Landau2020} found that resting-state EEG recordings could predict meaningful personality traits and cognitive abilities.

\begin{figure*}[htpb]\centering
{\includegraphics[width=0.8\linewidth,clip]{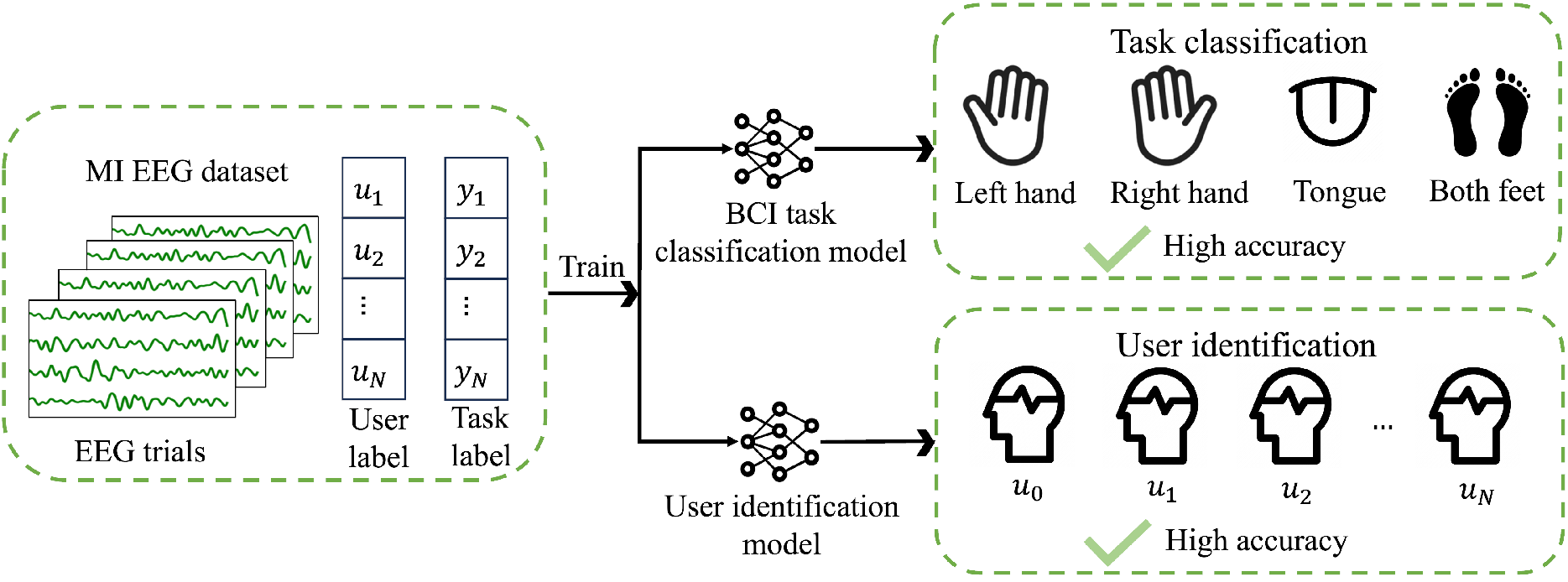}}
\caption{Both task classification model and UID model can be trained on the MI EEG dataset. The training of task classification model requires EEG trails and task labels (left hand, right hand, tongue, and both feet), and the training of UID model needs EEG trails and user labels.}
\label{fig:train}
\end{figure*}

With the development of BCI, an increasing number of public EEG datasets have been released.  These datasets often label which EEG data comes from the same user, facilitating the development of within-user or cross-user EEG task algorithms since EEG data varies significantly between users \cite{Wu2022}. However, this practice also raises privacy concerns. Machine learning models can learn user identity information from EEG data, enabling user identification (UID) by recognizing users who appear in public EEG datasets \cite{Zhang2022, Meng2023}. For instance, in MI-based BCIs (distinguishing imagined movements of the left hand, right hand, both feet, and tongue from EEGs), both task classification models and UID models can be trained on the same dataset \cite{Meng2023}, as shown in Figure~\ref{fig:train}. Once user identity information is learned from public EEG data, users can be identified through EEG. Private information such as emotional states \cite{Wu2023} and disease conditions \cite{Zhao2023} contained in EEG data may also be at risk of exposure.

Diverse privacy protection approaches have been proposed for EEG-based BCIs to prevent the leak of private information in EEG data, which can be broadly categorized into \cite{Xia2023}:
\begin{enumerate}
    \item Privacy-preserving machine learning, which avoids directly using raw EEG data or model parameters. Typical approaches include source-free transfer learning \cite{Xia2022, Zhang2023, Zhang2023}. Since the training data are not saved in the system, data privacy could be largely protected. For example, Zhang \emph{et al.} \cite{Zhang2022} studied gray-box and black-box model transfer for MI and affective BCI. Li \emph{et al.} \cite{Li2022} proposed multi-source model transfer with uninvertible training data in MI and ERP classification. Zhao \emph{et al.} \cite{Zhao2023} proposed source-free domain adaptation for privacy-preserving seizure subtype classification.% Jia \emph{et al.} \cite{Jia2024} eliminated the risk of data transfer from client users to the server model using federated learning.
    \item Synthetic data generation, which typically leverages generative models to synthesize data that retain the BCI task information. For example, Pascual \emph{et al.} \cite{Pascual2021} developed EpilepsyGAN to synthesize ictal seizure EEG signals, which retained task-related components and were challenging for UID. However, the performance of downstream models would be compromised if the fidelity of these synthesized data is low \cite{Xia2023}.
    \item Data perturbation, which adds perturbations to the original data to prevent machine learning models from learning private information while maintaining data utility for downstream tasks. Meng \emph{et al.} \cite{Meng2023} designed a sample-wise and a user-wise perturbation to generate identity-unlearnable EEG data. Sample-wise perturbation computes different perturbations for each EEG sample. User-wise perturbation calculates a different perturbation for each user and then add this perturbation to all of each user's EEG samples. User-wise perturbation is more computationally effective and have shown more promising performance \cite{Meng2023}.
\end{enumerate}
Privacy-preserving machine learning requires that user data is not saved in the system, which has special requirements for algorithm design. Synthetic data generation is very challenging; for example, generated adversarial networks (GAN) are prone to collapse and non-convergence during training. Data perturbation, a recently proposed new approach, is relatively simple and does not impose additional requirements on the algorithm design flow \cite{Huang2021}.

Huang \emph{et al.} \cite{Huang2021} revealed that adding perturbations to data can make them unlearnable for deep learning models in computer vision domain. Our previous work conducted by Meng \emph{et al.} \cite{Meng2023} revealed that adding data perturbation to EEG can make the deep learning model unable to distinguish which user the EEG data is from, i.e., unable to perform UID, and a good task classification model can be normally trained on the perturbed EEG data. However, the approach in our previous work is complicated, and the application scenario is limited, requiring adding perturbations to both training and test data. This raises concerns that their UID model may still learn user identity information from the training data, with perturbations on the test data preventing UID. In practical scenarios where test data is typically not perturbed, the UID model trained on perturbed datasets may still identify users, thereby leaking privacy information.

\begin{figure*}[htpb]\centering
{\includegraphics[width=0.9\linewidth,clip]{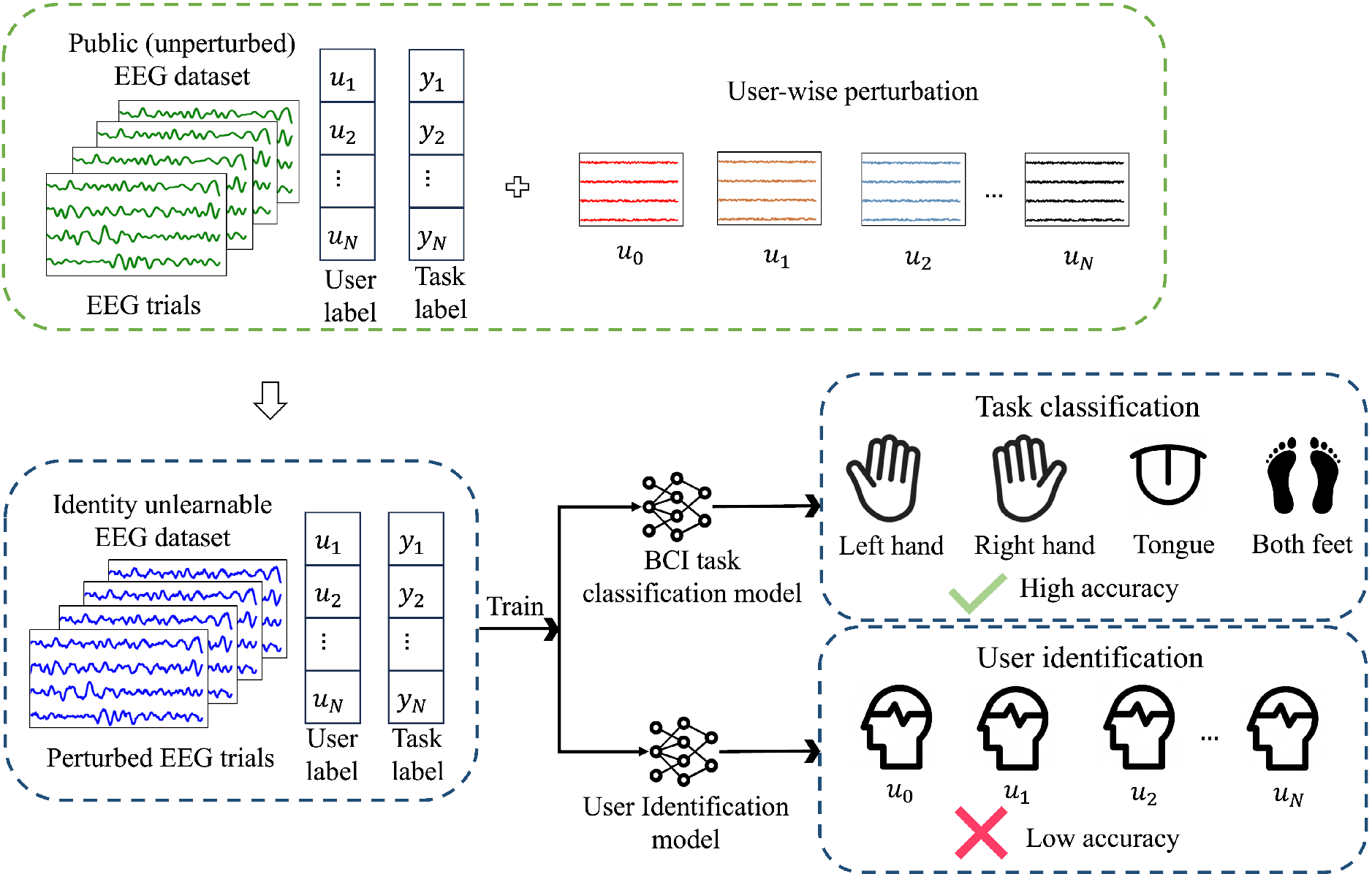}}
\caption{Illustration of user identity protection in EEG-based MI classification. The UID model trained on identity-unlearnable training EEG data cannot distinguish user identities from the original (unperturbed) test data. In contrast, the BCI task classifier trained on these identity-unlearnable EEG data still performs well on the original test data. This indicates that identity-unlearnable EEG data still contains learnable task information, but the user identity information is hidden and cannot be learned by the machine learning model.}
\label{fig:main}
\end{figure*}

 In this work, we demonstrate that adding user-wise perturbations to EEG data can effectively prevent models from learning the user identity information without significantly affecting the BCI task information of the EEG data for the first time. As illustrated in  Figure~\ref{fig:main}, a UID model trained on the perturbed data is unable to distinguish the user identity information on the original (unperturbed) test data, whereas the BCI task classifier trained on these perturbed training data has almost no performance degradation on the original test data, compared to the one trained on the unperturbed training data. In contrast to our previous study \cite{Meng2023}, which requires test data perturbation to prevent UID, our approach does not need any extra operations in the test phase, which ensures that the model can not learn identity information from the released public dataset (training EEG data).

Huang \emph{et al.} \cite{Huang2021} also showed that some methods, such as adversarial training (AT) and data transformation, may destroy the structure of privacy preserving perturbations and make deep learning models learn the private information in the data. We show experimentally that we design robust privacy preserving perturbations in EEG for the first time. We propose four types of user-wise perturbations for privacy protection and verify the robustness of our proposed perturbations under AT and data transformation. In summary, our main contributions are:
\begin{enumerate}
    \item  We show that user-wise perturbations can effectively make identity information in EEG data unlearnable for the first time, successfully protecting privacy information in EEG data without influencing task-related information.
    \item  We propose four types of user-wise perturbations for privacy protection in EEG: random noise (RAND), synthetic noise (SN), error minimization noise (EMIN), and error maximization noise (EMAX).
    \item  We empirically verify the effectiveness of the proposed user-wise privacy-preserving perturbations on deep learning models and traditional machine models. We also empirically verify their robustness under AT and multiple data transformation techniques, verifying their practicability in EEG experimentally for the first time.
\end{enumerate}

The remainder of this paper is organized as follows: Section~\ref{sect:met} introduces the proposed user-wise privacy-preserving perturbations. Section~\ref{sect:es} describes the experimental settings. Section~\ref{sect:er} presents the experimental results. Finally, Section~\ref{sect:CFR} draws conclusions and points out a future research direction.

\section{Methodology} \label{sect:met}

This section proposes four types of user-wise privacy-preserving perturbations, whose pseudo-code is shown in Algorithm~\ref{Alg:ppn}. The Python code is available at \href{https://github.com/xqchen914/Unlearnable-examples-for-EEG}{https://github.com/xqchen914/Unlearnable-examples-for-EEG}.

\begin{algorithm}[htpb]
    \caption{User-wise Privacy-Preserving Perturbations.}\label{Alg:ppn}
    \KwIn{$\mathcal{D}=\left\{\left(X_i, y_i, u_i\right)\right\}_{i=1}^N$, the original training EEG dataset\;
        \hspace*{10mm} $M$, the number of substitute models for UID\;}

    \KwOut{User-wise privacy-preserving perturbations RAND $\Delta^{RAND}_{u_i}$, SN $\Delta^{SN}_{u_i}$, EMIN $\Delta^{EMIN}_{u_i}$, and EMAX $\Delta^{EMAX}_{u_i}$.}
    \tcp{RAND}
    Generate $\Delta^{RAND}_{u_i}$ by (\ref{eq:rand})\;

    \tcp{SN}
    Generate $\Delta^{SN}_{u_i}$ by (\ref{eq:sn})\;

    \tcp{EMIN}
    Randomly initialize a UID model $D$\;
    Train $\Delta^{EMIN}_{u_i}$ using model $D$ and dataset $\mathcal{D}$ by (\ref{eq:emin}) and (\ref{eq:tanh})\;

    \tcp{EMAX}
    \For{$m=1, \cdots, M$}{
        Train UID model $D_m$ on $\mathcal{D}$\;
    }
    Train $\Delta^{EMAX}_{u_i}$ using $\{D_m\}_{m=1}^M$ and dataset $\mathcal{D}$ by (\ref{eq:tanh}) and (\ref{eq:emax}).
\end{algorithm}

\subsection{Problem Setup}

Given an EEG training dataset $\mathcal{D}=\left\{\left(X_i, y_i, u_i\right)\right\}_{i=1}^N$ comprising $N$ samples, where $X_i\in  \mathbb{R}^{c \times t}$ is the $i$-th EEG trial with $c$ channels and $t$ time domain samples, $y_i \in \{1, \ldots, K\}$ is the corresponding BCI task label (e.g., left/right hand MI), and $u_i \in \{1, \ldots, U\}$ is the corresponding user identity label. We use $D$ to stand for a UID model trained on dataset $\mathcal{D}$.

This paper aims to make the user identity information in EEG data unlearnable, while preserving the BCI task information, e.g., MI. We generate an identity-unlearnable EEG training dataset $\mathcal{D}^{\prime}=\left\{\left(X_i^{\prime}, y_i, u_i\right)\right\}_{i=1}^N$ from $\mathcal{D}$, by adding a user-wise perturbation $\Delta_{u_i}\in  \mathbb{R}^{c \times t}$ to each EEG trial $X_i$, i.e., $X_i^{\prime}=X_i + \Delta_{u_i}$.

\subsection{Random Noise (RAND)}

RAND uses user-specific random noise as the privacy-preserving perturbation, i.e.,
\begin{align}
    \Delta_{u_i}\leftarrow\alpha\cdot \mathrm{U}(-1,1),\label{eq:rand}
\end{align}
where $\alpha$ is a predefined perturbation amplitude, and $\mathrm{U}(-1,1)\in\mathbb{R}^{c \times t}$ is a matrix whose each element is a uniformly distributed number in $[-1,1]$. Note that $\Delta_{u_i}$ is the same for all EEG trials belonging to the same user but different for different users.

\subsection{Synthetic Noise (SN)}

For each user, SN selects a random integer, finds its 10-digit binary representation, replaces `0' in the binary representation with `$-1$', and then repeats each digit ten times to generate a square wave $\bm{w}_u$. For example, assume a random number 914 is chosen for the $i$-th subject; then, the 10-digit binary representation of 914 is $1110010010$, and $\bm{w}_{u_i}=\underbrace{1 \cdots 1}_{30} \underbrace{-1 \cdots-1}_{20} \underbrace{1 \cdots 1}_{10} \underbrace{-1 \cdots-1}_{20} \underbrace{1 \cdots 1}_{10} \underbrace{-1 \cdots-1}_{10}$.

Finally, the user-wise perturbation is:
\begin{align}
	\Delta_{u_i}=\alpha \cdot\begin{bmatrix}
    a_1 \\
    a_2 \\
    \vdots \\
    a_c
\end{bmatrix} \cdot \bm{w}_{u_i},\label{eq:sn}
\end{align}
where $\alpha$ is a predefined perturbation amplitude, and each $a_j$ ($j\in\{1,\cdots,c\}$) is a uniformly distributed random number in $[0.5, 1.5]$. For each user, the perturbation has the same waveform $\bm{w}_{u_i}$ for each EEG channel, but different magnitudes. If the length of $\bm{w}_{u_i}$ is unequal to $t$, the number of time domain samples of an EEG trial, then $\bm{w}_{u_i}$ is clipped or repeated to match $t$.

In SN, $\bm{w}_{u_i}$ adds perturbations to the time domain, whereas $\{a_1,\cdots,a_c\}$ adds perturbations to the spatial domain. In this way, SN fools the UID model to learn the perturbations rather than the user identity information.

\subsection{Error Minimization Noise (EMIN)}

EMIN optimizes user-wise perturbations $\Delta_{u_i}$ directly on a UID model $D$, by minimizing the cross-entropy loss $\mathcal{L}_{\mathrm{CE}}$ of the model prediction on the transformed data $D(\mathrm{trans}(X_i+\Delta_{u_i}))$ and the groundtruth label $u_i$, i.e.,
\begin{align}
\min _{\Delta_{u_i}}  \mathcal{L}_{\mathrm{CE}}\left(D(\mathrm{trans}(X_i+\Delta_{u_i})), u_i\right),\label{eq:emin}
\end{align}
where $\mathrm{trans}(\cdot)$ cuts an EEG trial into multiple time domain segments and then randomly concatenates them. This procedure would fool a UID model to learn the association between the randomly shuffled $\mathrm{trans}(X_i+\Delta_{u_i})$ and the user identity, instead of the user identity information, and hence protects the user identify information in $X_i$.

To keep the perturbation amplitude ($\ell_\infty$ norm) within a certain range, we introduce a new matrix $\Lambda_{u_i}\in \mathbb{R}^{c \times t}$: instead of optimizing $\Delta_{u_i}$, we optimize $\Lambda_{u_i}$ \cite{Carlini2017}. Specially, we set
\begin{align}
    \Delta_{u_i}=\alpha\cdot \tanh(\Lambda_{u_i}),\label{eq:tanh}
\end{align}
where $\alpha$ is a predefined perturbation amplitude. Since $\|\mathrm{tanh}(\Lambda_{u_i})\|_\infty<1$, we ensure that $\|\Delta_{u_i}\|_\infty<\alpha$.

\subsection{Error Maximization Noise (EMAX)}

EMAX, in contrast to EMIN, maximizes the cross-entropy loss $\mathcal{L}_{\mathrm{CE}}$ between the groundtruth user identity label $u_i$ and the predictions of $M$ substitute models $\{D_m\}_{m=1}^M$:
\begin{align}
\max _{\Delta_{u_i}} \sum_{m=1}^{M} \mathcal{L}_{\mathrm{CE}}\left(D_m(X_i+\Delta_{u_i}), u_i\right).\label{eq:emax}
\end{align}
Using multiple substitute models to optimize EMAX enhances the transferability of the perturbation \cite{Tramer2017}. We still use (\ref{eq:tanh}) to constrain the $\ell_\infty$ norm of $\Delta_{u_i}$. Similar to EMIN, we also optimize $\Lambda_{u_i}$ instead of $\Delta_{u_i}$ directly.

(\ref{eq:emax}) misleads the UID model to learn the noise patterns of the training data, which becomes incapable of distinguishing the true user identities on the original test data \cite{Fowl2021}.

\section{Experiment Settings} \label{sect:es}

This section introduces the datasets, models, evaluation metrics and hyperparameters in our experiments.

\subsection{Datasets}

The following six publicly available EEG datasets, summarized in Table~\ref{tab:data}, were used in this paper:

\begin{table*}[!t]
\scriptsize \centering \setlength{\tabcolsep}{1mm}
  \caption{Summary of the six datasets.}\label{tab:data}
    \begin{tabular}{c|c|c|c|c|c|c}
    \toprule
    Dataset & \# Subjects & \# Time Samples & \# Channels & \# Trials/Subject & \# Sessions & Classes \\
     \midrule
    MI1 & 9 & 1,000 & 22 & 144 & 2 & left hand, right hand, feet, tongue \\
    MI2 & 109 & 640 & 64 & 90 & 1 & left hand, right hand\\
    MI3 & 60 & 512 & 27 & 240 & 1 & left hand, right hand\\
    MI4 & 9 & 1,000 & 3 & 600 & 5 & left hand, right hand\\
    ERN & 16 & 166 & 56 & 340 & 5 & good feedback, bad feedback\\
    P300 & 10 & 206 & 16 & 1,728 & 3 & target, non-target \\
    \bottomrule
    \end{tabular}
\end{table*}

\begin{enumerate}
    \item Motor Imagery 1 (MI1) \cite{Tangermann2012} is Dataset 2a from BCI Competition IV. It includes data from nine subjects, each performing two sessions of four-class MI tasks on different days: left hand, right hand, feet, and tongue movements. EEG signals were recorded from 22 channels at a sampling rate of 250 Hz. After extracting data within [0, 4] seconds of each imagination prompt, we applied [8, 32] Hz band-pass filtering. Each subject had 144 EEG epochs per class.
    \item Motor imagery 2 (MI2) \cite{Schalk2004} includes EEG data from 109 users performing four tasks in 12 runs. Two tasks recorded in 6 runs, imagining the opening and closing of the left or right fist, were used in this paper. EEG signals were recorded from 64 channels at a sampling rate of 160 Hz. After extracting data within [0, 4] seconds of each imagination prompt, we applied [4, 32] Hz band-pass filtering. Each subject had 45 EEG epochs per class.
    \item Motor imagery 3 (MI3) \cite{Dreyer2023} includes EEG data from 60 users performing left hand and right hand MI tasks for 6 runs. EEG signals were recorded from 27 channels at a sampling rate of 512 Hz. After applying [4, 32] Hz band-pass filtering, we downsampled the data to 128 Hz and extracted data within [0, 4] seconds of each imagination prompt. Each subject had 120 epochs per class.
    \item  Motor imagery 4 (MI4) \cite{Leeb2007} includes EEG data from 9 users performing four tasks in five sessions. Two tasks, imagining the opening and closing of the left or right fist, were used in this paper. EEG signals were recorded from 3 channels at a sampling rate of 250 Hz. After extracting data within [0, 4] seconds of each imagination prompt, we applied [4, 32] Hz band-pass filtering. Each subject had around 300 EEG trials per class.
    \item Feedback error-related negativity (ERN) \cite{Soekadar2012} was employed in a competition during the 2015 IEEE Neural Engineering Conference on Kaggle. This paper utilized the training set, which included data from 16 users. Each user had 340 trials with two classes (good feedback and bad feedback) from five sessions. In preprocessing, the 56-channel EEG data was band-pass filtered within [1, 30] Hz and down-sampled to 128 Hz. The trial within [0, 1.3] seconds after each stimulus onset was extracted.
    \item P300 evoked potentials (P300) \cite{Hoffmann2008} were obtained from 10 users, each with 1728 EEG trials from two classes (target and non-target) across three sessions. In the preprocessing phase, trials within [0, 1] seconds after each image onset were extracted. The 16-channel EEG data were then detrended, band-pass filtered within [1, 40] Hz and down-sampled to 128 Hz.
\end{enumerate}

\subsection{Evaluation Metrics}

We used balanced classification accuracy (BCA) to evaluate the classification performance. The raw classification accuracy (RCA) is the ratio of the number of correctly classified examples to the number of total examples. The BCA is the average of the per-class RCAs. BCA was preferred in our experiments because the ERN and P300 datasets have significant intrinsic class imbalance, so using RCA is misleading. When all classes have the same number of samples, BCA reduces to RCA \cite{Wu2017, Zhang2019}.

In task classification, $\mathrm{BCA}=\frac{1}{K} \sum_{k=1}^K \frac{1}{N_k} \sum_{i=1}^{N_k} \mathbf{1}\left(y_{\text {pred }, i}=y_i\right)$, where $K$ denotes the number of classes, $N_k$ the number of test samples in class $k$, $y_{\text {pred }, i}$ the classifier's prediction on the $i$-th test sample, and $\mathbf{1}(\cdot)$ an indicator function. For UID, $\mathrm{BCA}=\frac{1}{U} \sum_{u=1}^U \frac{1}{N_u} \sum_{i=1}^{N_u} \mathbf{1}\left(u_{\text {pred }, i}=u_i\right)$, where $U$ is the number of users, $N_u$ the number of test samples belong to user $u$, $u_{\text {pred }, i}$ the classifier's prediction on the $i$-th test sample, and $u_i$ the true user identity.

\subsection{Classifiers}

Three popular CNN-based classifiers, i.e., EEGNet, DeepCNN, and ShallowCNN, were utilized in this paper for both UID and BCI task classification. Many studies have proved that these three networks have good EEG classification performance, and are suitable for various BCI paradigms \cite{Meng2023, Lawhern2018, Schirrmeister2017,Rajwal2023}.

\begin{enumerate}
    \item EEGNet \cite{Lawhern2018}. Designed explicitly for EEG classification tasks, EEGNet is a compact CNN architecture consisting of two convolutional blocks and a single classification block. It utilizes depthwise and separable convolutions, deviating from traditional convolutions, to effectively reduce the model's parameters.
    \item DeepCNN \cite{Schirrmeister2017}. Featuring a higher number of parameters compared to EEGNet, DeepCNN has three convolutional blocks and a softmax layer for classification. The initial convolutional block is tailored for EEG inputs, while the subsequent two blocks follow standard conventions.
    \item ShallowCNN \cite{Schirrmeister2017}. Derived from DeepCNN and inspired by filter bank common spatial patterns, ShallowCNN is a simplified variant. It includes a convolutional block with a larger kernel, a distinct activation function, and different pooling techniques compared to DeepCNN.
\end{enumerate}

We used the same network parameters as in \cite{Chen2024}.

\subsection{Experiment Settings and Hyperparameters}

We used the first 1/2/2/2/1/1 session(s)/block(s) in MI1/MI2/MI3/MI4/ERN/P300 datasets as the training data and the remaining session(s)/block(s) as the test data. All experiments were repeated five times and the average results are reported. This cross-session/block data division can effectively prevent the 'pitfall' problem in EEG recognition\cite{Li2021}, ensuring the rationality and effectiveness of the experiment.

The perturbation amplitude $\alpha$ was chosen as a multiple of the standard deviation of the user data. The multiples for RAND, SN, EMIN and EMAX on the six datasets are shown in Table~\ref{tab:hp}. $M=3$ substitute models were used in EMAX. When calculating the EMIN and EMAX perturbations, we performed gradient descent using the Adam optimizer for 100 epochs, using an EEGNet substitute model. Intuitively, the larger the amplitude $\alpha$ of the perturbation, the better the robustness of the privacy-preserving perturbation. However, it may have a more noticeable impact on the task information of the data. Nonetheless, within a certain amplitude range, the effects of privacy-preserving perturbations remained stable and robust, which was discussed in subsection \ref{sect:psa}.

\begin{table*}[htbp] \footnotesize  \centering \setlength{\tabcolsep}{5mm}
  \caption{The user-wise privacy-preserving perturbation amplitude $\alpha$.}\label{tab:hp}
    \begin{tabular}{c|cccc}
    \toprule
    Dataset & RAND & SN & EMIN & EMAX \\
    \midrule
    MI1 & 0.5 & 0.5 & 0.3 & 0.3 \\
    MI2 & 0.5 & 0.5 & 0.3 & 0.3 \\
    MI3 & 0.5 & 0.5 & 0.3 & 0.3 \\
    MI4 & 1.0 & 1.0 & 1.0 & 1.0 \\
    ERN & 1.0 & 1.0 & 0.5 & 0.5 \\
    P300 & 1.0 & 1.0 & 1.0 & 1.0 \\
    \bottomrule
    \end{tabular}%
\end{table*}%

All neural network models were trained for 100 epochs with initial learning rate 0.01, which was reduced to 0.001 after 50 epochs. Batch size 128 was used on all datasets.

\section{Results and Analyses} \label{sect:er}

This section presents our experiment results.

\subsection{Main Results} \label{sect:ma}

We trained the BCI task classifiers and the UID models on the original (unperturbed) training data, as well as their various perturbed counterparts, and tested the models on the original (unperturbed) test data. The results are shown in Table~\ref{tab:main}:
\begin{enumerate}
    \item Because the original EEG data contains BCI task and user identity information, good BCI task classifiers and UID models can be trained on the same EEG dataset. Generally, better data quality led to better classification performance for both models.
    \item The BCAs of the three networks were similar in EEG task classification, but the BCAs of UID were very different. It may be that the user identity and task information in EEG data exist in different aspects, and the performance of EEG UID is more sensitive to these three different network structures.
    \item User-wise perturbations can prevent the UID model from learning user identity information in EEG datasets. After the UID models were trained on the perturbed data, their BCAs on the original test EEG samples were close to random. Note again that our user-wise perturbation approach does not modify the test data, which is much more convenient to implement than our previous approach \cite{Meng2023} and ensures that models can not learn user identity information from training EEG samples.
    \item The proposed four types of user-wise perturbations, i.e., RAND, SN, EMIN and EMAX, effectively hid user identity information in the training EEG data. Overall, SN, EMIN and EMAX outperformed RAND (i.e., the test BCAs of the corresponding UID models were smaller), indicating the effectiveness of more sophisticated perturbations over random noise.
    \item BCAs of BCI task classifiers trained on perturbed EEG data were similar to those trained on original data, indicating no significant impact of perturbations on the BCI task information. However, adding perturbations may increase the distribution bias between training and test data, usually resulting in slightly lower BCAs for models trained on perturbed data. An exception was observed with P300 data, where the BCA of EEGNet trained on data with RAND perturbation was higher, suggesting that the perturbations served as data augmentation.
\end{enumerate}

\begin{table*}[htbp] \scriptsize   \centering \setlength{\tabcolsep}{0.8mm}
  \caption{Test BCAs of BCI task classifiers and UID models trained on the original (unperturbed) EEG data and those with user-wise perturbations.} \label{tab:main}
    \begin{tabular}{c|c|c|cc|cc|cc|cc|cc|cc}
    \toprule
    \multirow{2}{*}{Dataset}& \multirow{2}{*}{\# of users} & \multirow{2}{*}{Model} & \multicolumn{2}{c|}{Original} & \multicolumn{2}{c|}{RAND} & \multicolumn{2}{c|}{SN} & \multicolumn{2}{c|}{EMIN} & \multicolumn{2}{c|}{EMAX} &   \multicolumn{2}{c}{Average Reduction}\\% \cline{4-17}
          &  &  & Task  & UID   & Task  & UID   & Task  & UID   & Task  & UID   & Task  & UID  & Task  & UID \\
    \midrule
    \multirow{3}{*}{MI1}& \multirow{3}{*}{9} & EEGNet & 61.85 & 87.93 & 61.67 & 11.28 & 60.82 & 11.44 & 60.81 & 11.12 & 60.65 & 12.61 & 0.86 & 76.32\\
&& DeepCNN & 63.82 & 92.79 & 62.45 & 11.48 & 61.77 & 12.00 & 62.64 & 10.73 & 61.91 & 9.02 & 1.63 & 81.98\\
&& ShallowCNN & 66.10 & 99.02 & 65.25 & 11.17 & 64.18 & 13.19 & 64.41 & 11.13 & 63.80 & 14.25 & 1.69 & 86.58\\
    \midrule
    \multirow{3}{*}{MI2}& \multirow{3}{*}{109} & EEGNet & 71.07 & 53.99 & 71.06 & 1.28  & 69.84 & 2.31  & 69.19 & 0.96  & 69.93 & 0.94 & 1.07 & 52.62\\
&& DeepCNN & 72.76 & 72.46 & 71.57 & 1.57  & 72.13 & 2.21  & 71.80 & 1.53  & 71.87 & 1.62 &  0.92 & 70.73\\
&& ShallowCNN & 72.71 & 87.39 & 71.19 & 6.91  & 69.23 & 3.83  & 70.11 & 1.16  & 70.29 & 3.98 & 2.51 & 83.42\\
    \midrule
    \multirow{3}{*}{MI3}& \multirow{3}{*}{60} & EEGNet &  81.70 & 66.10 & 79.52 & 2.28 & 79.33 & 2.37 & 78.20 & 1.70 & 79.82 & 1.70 & 2.48 & 64.09\\
&& DeepCNN &  83.59 & 87.56 & 83.00 & 2.04 & 82.03 & 2.88 & 80.60 & 1.95 & 83.07 & 1.39 & 1.42 & 85.50\\
&& ShallowCNN &  80.33 & 89.76 & 78.41 & 2.21 & 79.93 & 2.92 & 77.57 & 1.61 & 79.09 & 2.48 & 1.58 & 87.46\\
    \midrule
    \multirow{3}{*}{MI4}& \multirow{3}{*}{9} & EEGNet & 74.24 & 75.50 & 73.90 & 11.26 & 73.46 & 10.75 & 73.73 & 11.17 & 73.83 & 9.52 & 0.51 & 64.83\\
&& DeepCNN & 73.27 & 76.09 & 73.05 & 12.06 & 72.80 & 11.19 & 72.23 & 11.80 & 72.84 & 10.27 & 0.54 & 64.76\\
&& ShallowCNN & 73.58 & 79.71 & 73.31 & 14.89 & 72.16 & 12.73 & 72.69 & 11.35 & 72.35 & 9.99 & 0.95 & 67.47\\
    \midrule
    \multirow{3}{*}{ERN}& \multirow{3}{*}{16} & EEGNet & 65.26 & 41.43 & 63.94 & 6.08  & 63.82 & 6.22  & 63.62 & 6.52 & 64.32 & 5.54 & 1.34 & 35.34\\
&&  DeepCNN & 64.11 & 65.96 & 61.34 & 8.06  & 63.37 & 6.11  &  61.13 & 6.54 & 62.09 & 5.02 & 2.13 & 59.52 \\
&& ShallowCNN  & 65.19 & 45.44 & 64.76 & 9.79 & 63.85 & 10.13 &   64.90 & 9.10 & 64.23 & 11.73 & 0.76 & 35.25\\
    \midrule
    \multirow{3}{*}{P300}& \multirow{3}{*}{10} & EEGNet &   82.12 & 85.81 & 85.86 & 11.42 & 80.49 & 10.18 & 80.42 & 9.32 & 81.45 & 10.00 & 0.06 & 75.58\\
&& DeepCNN &  82.43 & 96.99 & 81.14 & 11.36 & 81.38 & 11.13 & 80.59 & 9.58 & 81.74 & 9.00 & 1.22 & 86.72\\
&& ShallowCNN &  81.91 & 96.69 & 81.28 & 13.01 & 80.92 & 14.27 & 80.91 & 12.52 & 81.71 & 13.64 & 0.71 & 83.33\\
    \midrule
    \multicolumn{3}{c|}{Average} &68.99 & 76.35 & 67.94 & 12.58 & 67.35 & 11.13 & 67.76 & 9.14  & 67.82 & 10.96& 1.27 & 65.40\\
    \bottomrule
    \end{tabular}
\end{table*}

\begin{figure*}[h]\centering
\subfigure[]{\includegraphics[width=0.49\columnwidth,clip]{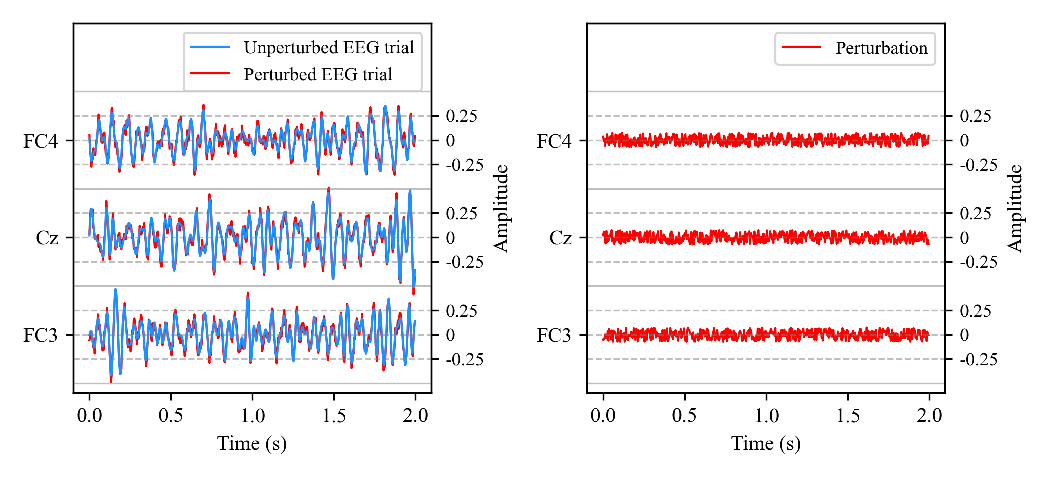}}
\subfigure[]{\includegraphics[width=0.49\columnwidth,clip]{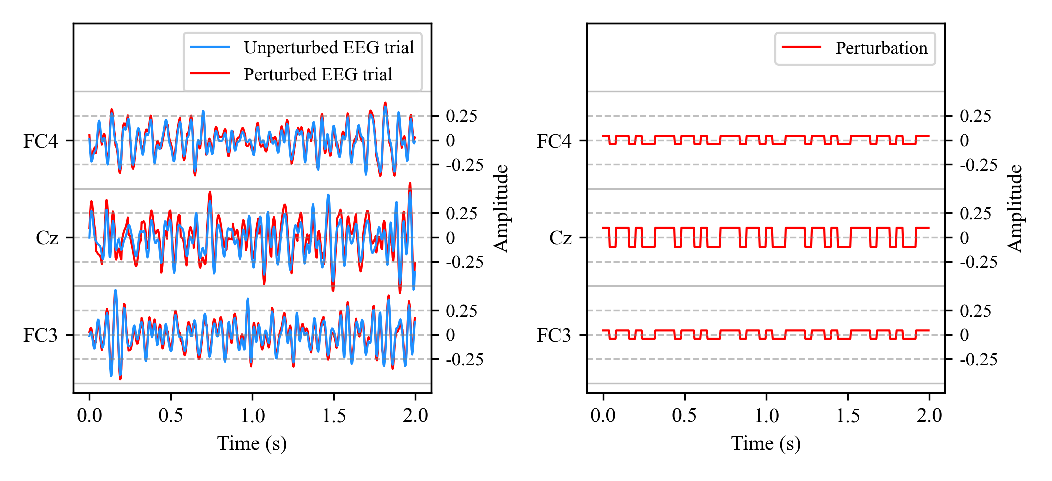}}
\subfigure[]{\includegraphics[width=0.49\linewidth,clip]{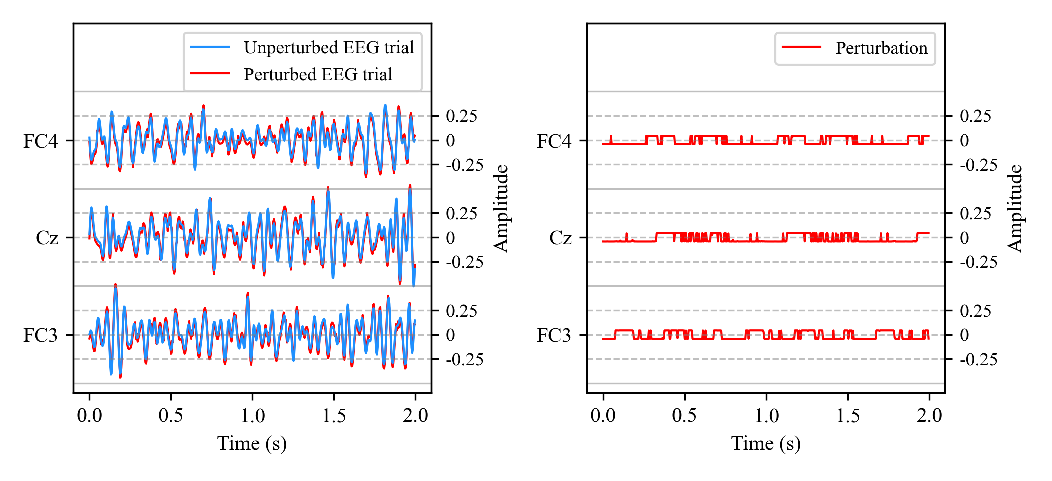}}
\subfigure[]{\includegraphics[width=0.49\linewidth,clip]{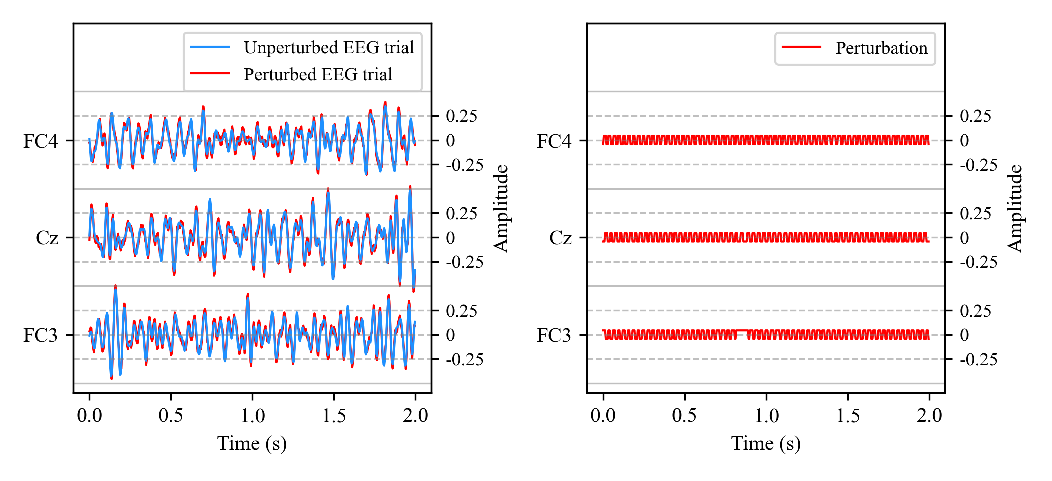}}
\caption{An original EEG trial and its perturbed counterpart with (a) RAND, (b) SN, (c) EMIN, and (d) EMAX.}\label{fig:perturb}
\end{figure*}

Figure~\ref{fig:perturb} shows that the original EEG trial and its perturbed counterparts were very similar, explaining why the BCI task classifiers had almost no performance degradations. We further calculated the normalized cross-correlation \cite{Stojanovic2006} between the original EEG trials and those with added RAND/SN/EMIN/EMAX perturbations on MI1, which was 0.96, 0.88, 0.96, and 0.96, respectively (the cross-correlation between EEG trials and themselves was always 1). This indicates that the original EEG trials are indeed highly similar to those with privacy-preserving perturbations. Moreover, as shown in Figure~\ref{fig:perturb}, the four approaches exhibited significantly different perturbation patterns.

In summary, all four proposed user-wise perturbations can significantly reduce the UID BCA (i.e., hide user identity information), with little sacrifice in the BCI task classification performance. Particularly, the more sophisticated EMIN and EMAX can hide more user identity information with generally a smaller perturbation amplitude.

\subsection{Robustness under AT}

Recent studies have discovered that EEG-based BCIs are vulnerable to adversarial attacks \cite{Zhang2019, Zhang2021}, where carefully designed tiny adversarial perturbations added to EEG trials can mislead classifiers into producing incorrect outputs. AT, which involves augmenting the training set with adversarial examples, is an effective approach to help models learn robust features from data and defend against adversarial attacks \cite{Meng2023a, Chen2024}. Thus, it is important to study if our proposed user-wise privacy-preserving perturbations can withstand AT, i.e., whether adversarial examples generated from our perturbed EEG data can break privacy-preserving perturbations and contain user identity information \cite{Huang2021}.

\begin{figure*}\centering
\subfigure[]{\includegraphics[width=.25\linewidth,clip]{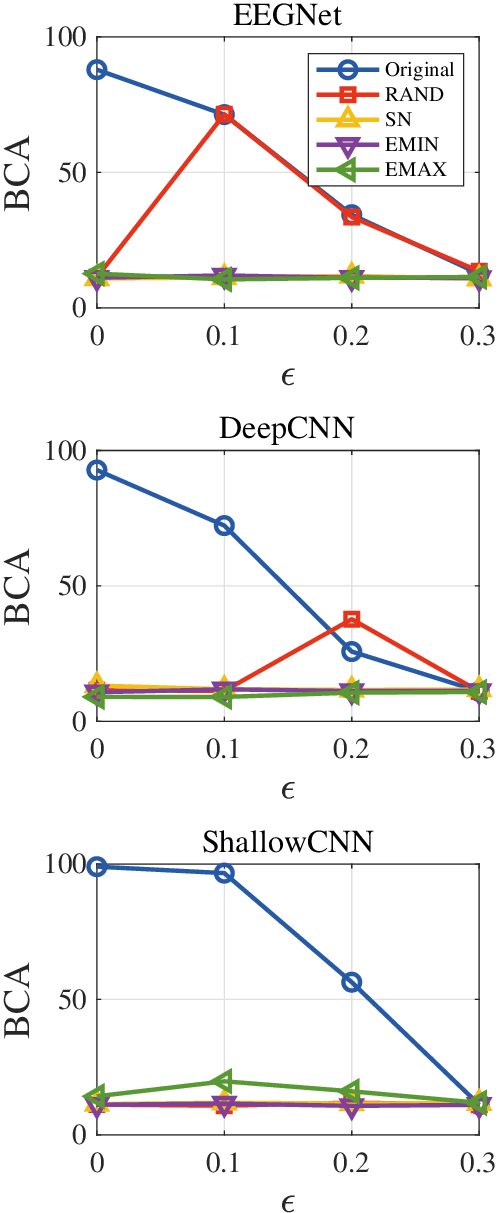}}
\subfigure[]{\includegraphics[width=.25\linewidth,clip]{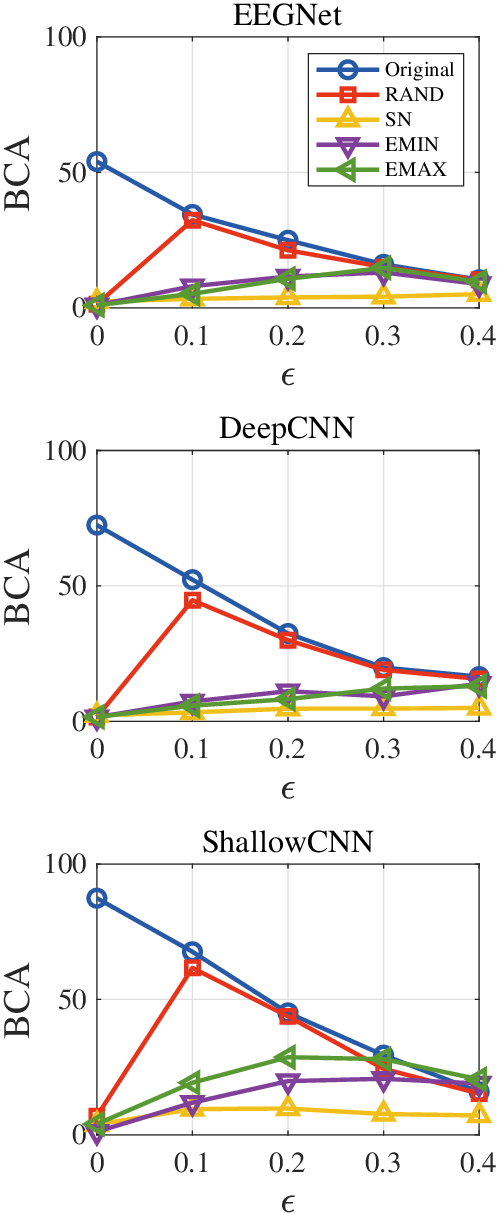}}
\subfigure[]{\includegraphics[width=.25\linewidth,clip]{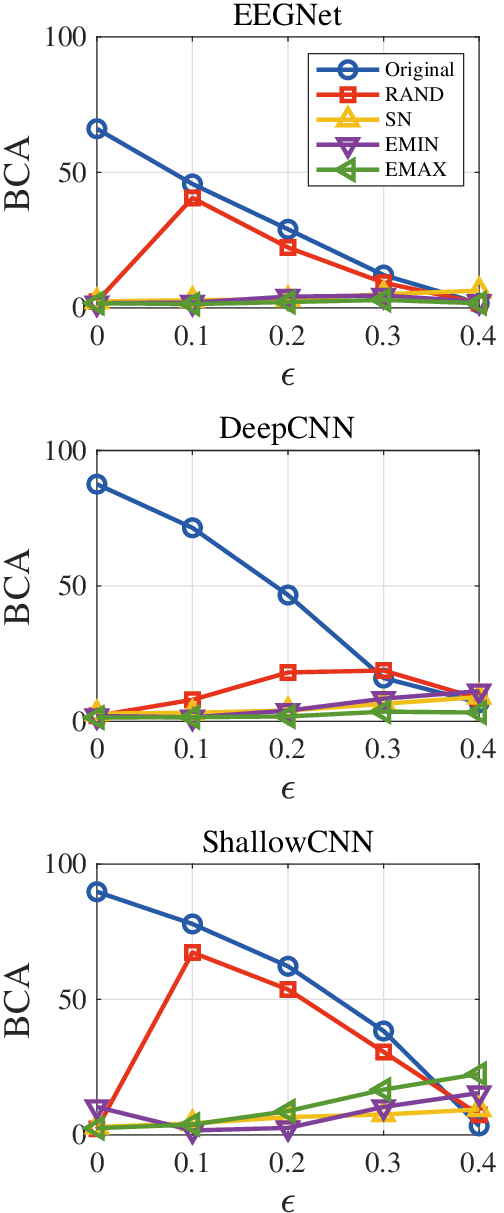}}
\subfigure[]{\includegraphics[width=.25\linewidth,clip]{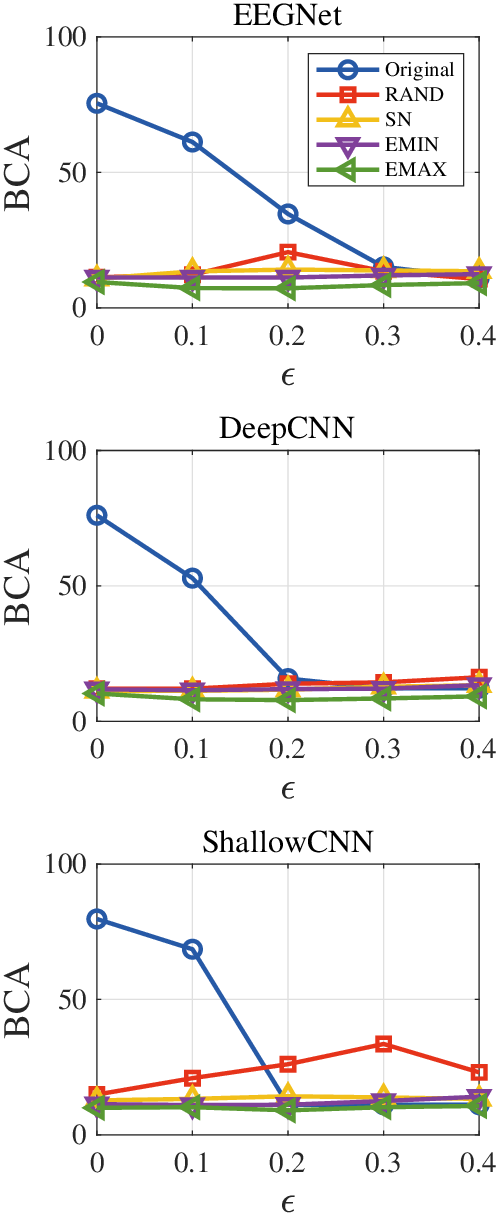}}
\subfigure[]{\includegraphics[width=.25\linewidth,clip]{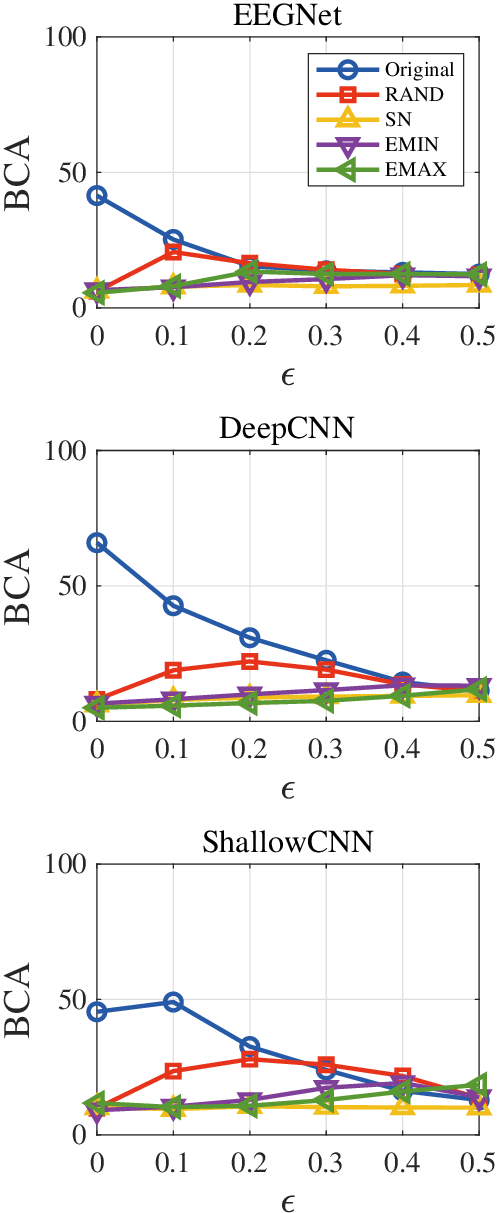}}
\subfigure[]{\includegraphics[width=.25\linewidth,clip]{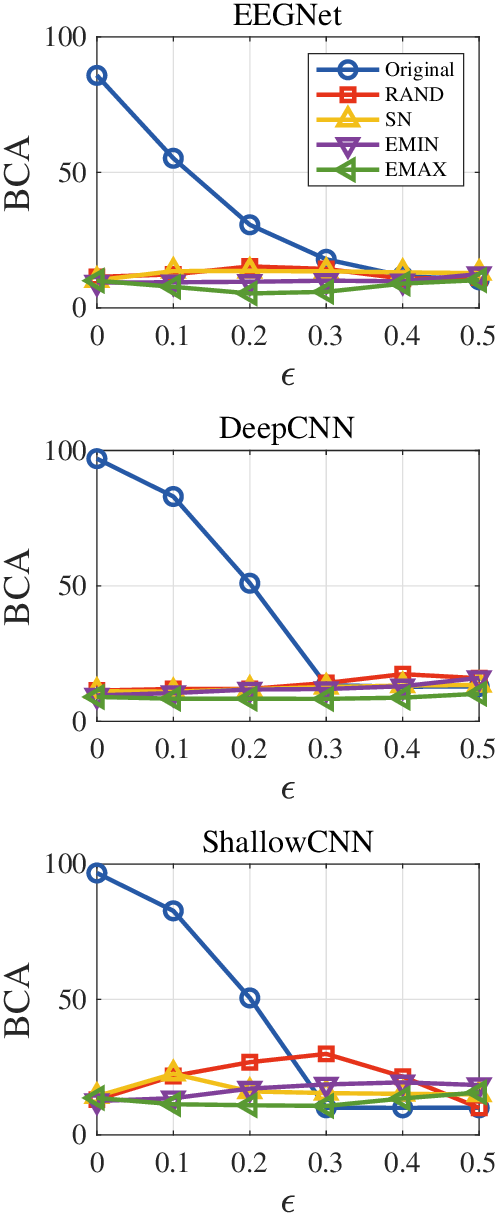}}
\caption{Test BCAs of different UID models under AT on (a) MI1; (b) MI2; (c) MI3; (d) MI4; (e) ERN; and (f) P300.}\label{fig:at}
\end{figure*}

Figure~\ref{fig:at} shows the results, where the adversarial examples were generated by the popular projected gradient descent \cite{Madry2018} approach. The horizontal axis indicates the adversarial perturbation amplitude $\epsilon$ used in projected gradient descent. A larger $\epsilon$ corresponds to a larger distortion applied to EEG data during AT. Observe that:
\begin{enumerate}
	\item In AT, when the amplitude $\epsilon$ of adversarial perturbation increased, the robustness of the model was improved, but its accuracy on benign samples generally decreased \cite{Meng2023a, Li2022}. As the amplitude of the adversarial perturbation increased, BCAs of UID models trained on the original data decreased.
	\item  On the data with RAND perturbation, with the increase of the amplitude $\epsilon$ of adversarial perturbation, the BCAs of the UID models increased first and then decreased, indicating that AT weakened the effect of RAND privacy protection to a certain extent. However, as the amplitude of adversarial perturbation further increased, the BCAs of UID models eventually decreased. This was particularly evident on the MI2 dataset.
	\item The three more sophisticated approaches (SN, EMIN and EMAX) demonstrated robustness (low UID BCA) across different datasets, classifiers, and adversarial perturbation amplitudes, indicating that these methods can effectively protect user identity information even under AT.
\end{enumerate}

\subsection{Robustness under Data Preprocessing/Transformation}

Huang \emph{et al.} \cite{Huang2021} found that various data transformation methods in the image analysis would change the structure of privacy-preserving perturbations and weaken their effect. It is also important to investigate whether data transformation approaches can impact the effect of privacy-preserving perturbations in EEG.

We considered the following data preprocessing/transformation approaches, which can be used to enhance MI classification performance:
\begin{enumerate}
    \item Surface Laplacian \cite{Carvalhaes2015}, which applies surface Laplacian to EEG data.
    \item Temporal Shift \cite{Meng2023a}, which shifts an EEG trial left or right with a random offset.
    \item Temporal Recombination, which cuts each EEG trial into multiple segments in the time domain, and then randomly shuffles and concatenates them.
\end{enumerate}

\begin{figure*}\centering
\subfigure[]{\includegraphics[width=.24\linewidth,clip]{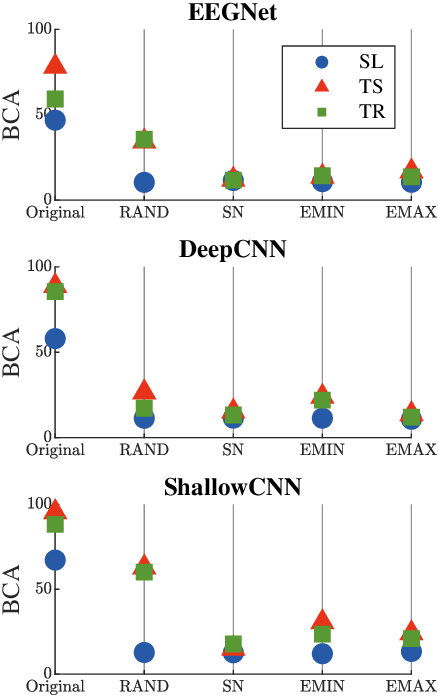}}
\subfigure[]{\includegraphics[width=.24\linewidth,clip]{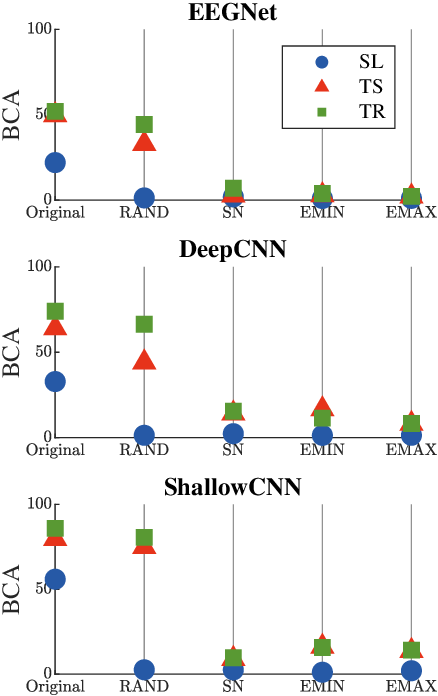}}
\subfigure[]{\includegraphics[width=.24\linewidth,clip]{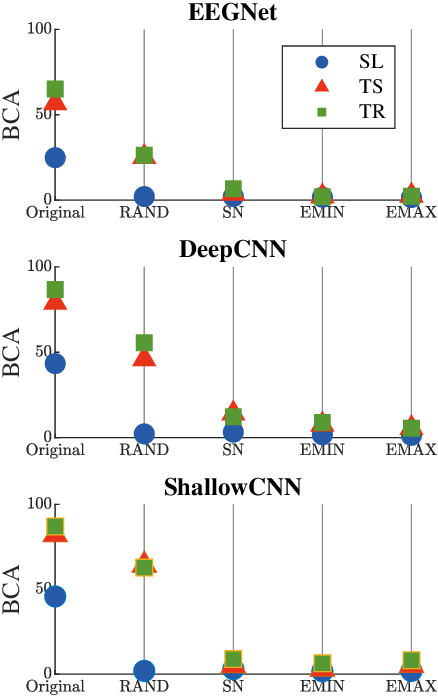}}
\subfigure[]{\includegraphics[width=.24\linewidth,clip]{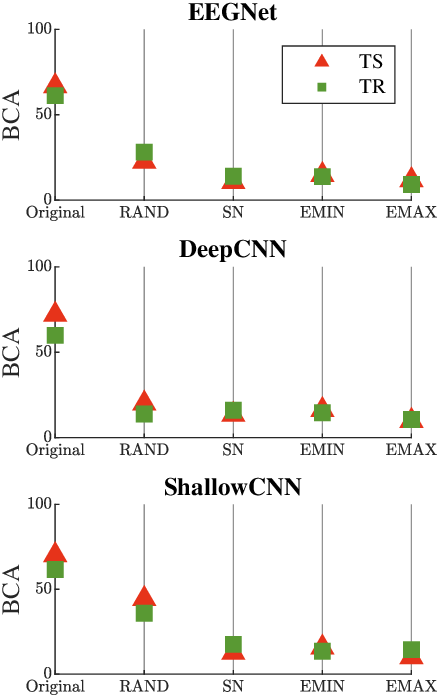}}
\caption{Test BCAs of the UID models under different data preprocessing/transformations on (a) MI1, (b) MI2, (c) MI3, and (d) MI4. SL: Surface Laplacian; TS: Temporal Shift; TR: Temporal Recombination. SL, which needs at least four channels, was not performed on MI4, as it has only three channels.}\label{fig:trans}
\end{figure*}

Figure~\ref{fig:trans} shows the results on the four MI datasets and three neural networks. Observe that:
\begin{enumerate}
	\item After data transformation, the UID BCAs of models on the original EEG sometimes decreased. This suggests that some identity information in the EEG may be embedded in the temporal or spatial structures of the original EEG data.
	\item The privacy-protection ability of RAND was significantly weakened (higher UID BCA were observed) under temporal shift/recombination, especially on MI2 and MI3 datasets. This indicates that RAND has relatively weak robustness, as simple data transformations can disrupt its structure, allowing the model to bypass RAND privacy-preserving perturbations and learn user privacy information from the data.
	\item SN, EMIN and EMAX were robust to different data preprocessing/transformations. On four datasets and three neural network models, the UID BCAs remained low under three data transformations when SN, EMIN or EMAX privacy-preserving perturbations were applied.
\end{enumerate}

\subsection{Applicability to Traditional UID Models}

Deep neural networks are trained through multiple rounds of gradient descent, whereas traditional machine learning models depend on expert knowledge. These two approaches have fundamentally different characteristics. Privacy-preserving perturbations exploit the vulnerabilities of deep neural networks \cite{Huang2021, Meng2023}, but their effectiveness on traditional machine learning models has yet to be studied. This subsection tests the effect of privacy-preserving perturbations on traditional UID models, e.g., Wavelet packet decomposition (Wavelet) \cite{Gui2014}, short-time Fourier transform (STFT) \cite{ArnauGonzalez2021}, and autoregressive (AR) \cite{Campisi2011}. After feature extraction, XGBoost \cite{Fowl2021} and linear discriminant analysis (LDA) \cite{Zhang2021a} were used for classification.

Table~\ref{tab:uid_tra} shows the UID results on the six datasets. Observe that:
\begin{enumerate}
	\item Although the UID BCAs were lower than those of the deep learning models in Table~\ref{tab:main}, traditional machine learning models still demonstrated good UID performance on the original EEG trials. LDA outperformed XGBoost, achieving a higher average UID BCA on the original EEG trials.
	\item Our proposed four types of user-wise perturbations can also prevent these traditional UID models from learning UID information. After adding privacy-preserving perturbations, the UID BCAs of various traditional machine learning models remained very low, indicating the robustness and broad applicability of our proposed privacy-preserving perturbations.
\end{enumerate}

There are some well-performing traditional UID machine learning models \cite{Zhang2021a}. In the future, privacy-preserving perturbations may also be designed from the perspective of traditional machine learning models. For instance, perturbations could be applied to manually designed features for privacy protection.

\begin{table}[htbp]
  \centering
  \setlength{\tabcolsep}{0.8mm}
  \scriptsize
  \caption{Test BCAs of traditional UID models trained on the original and perturbed training set.}
    \begin{tabular}{cc|ccccc|ccccc|c}
    \toprule
    \multicolumn{1}{c|}{\multirow{2}{*}{Dataset}} & \multirow{2}{*}{Feature} & \multicolumn{5}{c|}{Xgboost} & \multicolumn{5}{c|}{LDA} & \multirow{2}{*}{Average Reduction} \\
    \multicolumn{1}{c|}{} &   & Original & RAND & SN & EMIN & EMAX & Original & RAND & SN & EMIN & EMAX &  \\
    \midrule
    \multicolumn{1}{c|}{\multirow{3}[2]{*}{MI1}} & Wavelet & 49.77 & 11.11 & 11.11 & 11.11 & 11.11 & 49.69 & 11.11 & 11.11 & 11.11 & 11.11 & 38.62 \\
    \multicolumn{1}{c|}{} & STFT & 48.77 & 13.34 & 10.62 & 11.13 & 13.79 & 44.44 & 11.11 & 11.11 & 11.11 & 11.11 & 34.94 \\
    \multicolumn{1}{c|}{} & AR & 81.67 & 11.11 & 11.11 & 11.11 & 11.11 & 96.88 & 11.11 & 11.11 & 11.14 & 11.11 & 78.16 \\
    \midrule
    \multicolumn{1}{c|}{\multirow{3}[2]{*}{MI2}} & Wavelet & 63.31 & 3.93 & 12.86 & 5.56 & 3.81 & 87.20 & 0.92 & 2.27 & 1.03 & 0.94 & 71.34 \\
    \multicolumn{1}{c|}{} & STFT & 65.47 & 23.21 & 10.21 & 3.53 & 7.66 & 74.25 & 25.09 & 2.70 & 8.24 & 9.95 & 58.54 \\
    \multicolumn{1}{c|}{} & AR & 42.08 & 0.92 & 0.92 & 0.92 & 0.92 & 83.14 & 0.95 & 0.83 & 0.98 & 0.98 & 61.68 \\
    \midrule
    \multicolumn{1}{c|}{\multirow{3}[1]{*}{MI3}} & Wavelet & 72.03 & 5.36 & 6.74 & 2.92 & 1.70 & 89.44 & 2.01 & 3.15 & 1.77 & 1.64 & 77.57 \\
    \multicolumn{1}{c|}{} & STFT & 70.19 & 2.18 & 2.99 & 1.95 & 2.31 & 76.70 & 2.42 & 1.97 & 1.68 & 3.12 & 71.12 \\
    \multicolumn{1}{c|}{} & AR & 43.17 & 1.64 & 1.67 & 1.67 & 1.91 & 84.65 & 1.66 & 3.46 & 1.58 & 2.32 & 61.92 \\
    \midrule
    \multicolumn{1}{c|}{\multirow{3}[1]{*}{MI4}} & Wavelet & 71.41 & 11.06 & 11.31 & 11.08 & 11.44 & 74.11 & 11.11 & 11.11 & 11.11 & 11.11 & 61.59 \\
    \multicolumn{1}{c|}{} & STFT & 74.95 & 16.15 & 11.12 & 11.25 & 11.72 & 62.55 & 12.49 & 11.38 & 11.08 & 12.40 & 56.55 \\
    \multicolumn{1}{c|}{} & AR & 53.54 & 11.11 & 11.11 & 11.11 & 11.11 & 79.63 & 11.11 & 11.11 & 11.11 & 11.11 & 55.48 \\
    \midrule
    \multicolumn{1}{c|}{\multirow{3}[2]{*}{ERN}} & Wavelet & 39.15 & 3.96 & 10.33 & 7.04 & 6.25 & 51.61 & 6.25 & 6.46 & 6.21 & 5.97 & 38.82 \\
    \multicolumn{1}{c|}{} & STFT & 46.09 & 5.44 & 12.50 & 10.25 & 5.47 & 70.71 & 6.29 & 6.28 & 7.12 & 13.31 & 50.07 \\
    \multicolumn{1}{c|}{} & AR & 46.61 & 6.03 & 8.06 & 7.47 & 6.22 & 54.71 & 4.85 & 10.07 & 6.01 & 6.64 & 43.74 \\
    \midrule
    \multicolumn{1}{c|}{\multirow{3}[1]{*}{P300}} & Wavelet & 66.52 & 9.84 & 9.92 & 10.39 & 9.93 & 80.70 & 10.00 & 9.98 & 10.00 & 10.00 & 63.60 \\
    \multicolumn{1}{c|}{} & STFT & 68.82 & 9.98 & 10.05 & 10.30 & 10.03 & 75.49 & 11.36 & 10.05 & 10.05 & 10.34 & 61.89 \\
    \multicolumn{1}{c|}{} & AR & 60.63 & 10.00 & 10.00 & 10.00 & 10.00 & 77.21 & 10.00 & 9.97 & 10.00 & 10.00 & 58.92 \\
    \midrule
    \multicolumn{2}{c|}{Average} & 59.12 & 8.69 & 9.04 & 7.71 & 7.58 & 72.95 & 8.32 & 7.45 & 7.30 & 7.95 & 58.03 \\
    \bottomrule
    \end{tabular}%
  \label{tab:uid_tra}%
\end{table}%

\subsection{Ablation Analyses} \label{sect:abla}

We used different amplitudes for each EEG channel (channel variation) in SN, $\mathrm{trans(\cdot)}$ transformation in EMIN, and an ensemble of multiple UID models (model ensemble) in EMAX. This subsection conducts ablation analyses to evaluate how these strategies contributed to the robustness of the corresponding privacy-preserving perturbations.

The test BCAs of different UID models under AT and trained with different data transformation techniques on MI1, MI2 and P300 are shown in Tables~\ref{tab:ablaMI1}-\ref{tab:ablaP300}, respectively. Observe that:
\begin{enumerate}
    \item Channel variation contributed significantly to the robustness of SN. On all three datasets, when channel variation was not employed, the test BCAs of both ShallowCNN and DeepCNN UID models were much higher (i.e., much worse user identity protection ability) under surface Laplacian, temporal shift and temporal recombination.
    \item On MI1 and MI2, the $\mathrm{trans(\cdot)}$ transformation significantly improved the robustness of EMIN, especially under temporal shift/recombination. Without $\mathrm{trans(\cdot)}$, the privacy protection effectiveness of EMIN was substantially reduced (with a notable increase in the models' UID BCAs).
    \item EMAX showed good robustness in most cases without model ensemble, but not when ShallowCNN was used on MI2, indicating that model ensemble improves the robustness of EMAX in more situations.
    \item Compared to data transformation approaches, AT was generally less effective in disrupting privacy-preserving perturbations. Three types of privacy-preserving perturbations were still significantly effective under AT (with low UID BCAs of models).
\end{enumerate}
In summary, the tricks in SN, EMIN and EMAX are all necessary, as each of them enhanced the robustness of the corresponding privacy-prevserving perturbations in certain cases.

\begin{table*}[htbp] \footnotesize   \centering \setlength{\tabcolsep}{1.3mm}
\tabcolsep 3pt
  \caption{Test BCAs of different UID models in ablation analysis on MI1.}\label{tab:ablaMI1}
     \begin{tabular}{c|c|c|c|c|c|c}
    \toprule
     \multirow{2}{*}{Approach} & \multirow{2}{*}{Model} & \multirow{2}{*}{Setting} & Adversarial & Surface & Temporal  & Temporal \\
&  & & Training & Laplacian & Shift  & Recombination \\
    \midrule
    \multirow{6}[6]{*}{SN} & \multirow{2}[2]{*}{EEGNet} & w/o channel variation & 10.87 & 48.81 & 16.81 & 23.23 \\
      &   & w/ channel variation & 11.19 & 11.53 & 11.99 & 11.73 \\
\cmidrule{2-7}      & \multirow{2}[2]{*}{DeepCNN} & w/o channel variation & 11.81 & 60.36 & 36.99 & 45.60 \\
      &   & w/ channel variation & 11.85 & 11.44 & 15.29& 13.29 \\
\cmidrule{2-7}      & \multirow{2}[2]{*}{ShallowCNN} & w/o channel variation & 11.34 & 67.66 & 32.14 & 35.39 \\
      &   & w/ channel variation & 11.85 & 12.57 & 15.01  & 17.79 \\
    \midrule
    \multirow{6}[6]{*}{EMIN} & \multirow{2}[2]{*}{EEGNet} & w/o $\mathrm{trans(\cdot)}$ & 10.73 & 11.30 & 23.81  & 25.36 \\
      &   & w/ $\mathrm{trans(\cdot)}$ & 11.94 & 10.83 & 13.79  & 14.27 \\
\cmidrule{2-7}      & \multirow{2}[2]{*}{DeepCNN} & w/o $\mathrm{trans(\cdot)}$ & 10.69 & 11.80 & 48.64  & 56.21 \\
      &   & w/ $\mathrm{trans(\cdot)}$ & 11.82 & 11.41 & 23.97 & 21.96 \\
\cmidrule{2-7}      & \multirow{2}[2]{*}{ShallowCNN} & w/o $\mathrm{trans(\cdot)}$ & 11.24 & 10.96 & 66.70  & 45.55 \\
      &   & w/ $\mathrm{trans(\cdot)}$ & 11.57 & 12.01 & 30.56  & 23.54 \\
    \midrule
    \multirow{6}[6]{*}{EMAX} & \multirow{2}[2]{*}{EEGNet} & w/o model ensemble & 10.94 & 9.93 & 15.73 & 18.03 \\
      &   & w/ model ensemble & 10.48 & 10.45 & 16.82  & 13.66 \\
\cmidrule{2-7}      & \multirow{2}[2]{*}{DeepCNN} & w/o model ensemble & 10.71 & 11.01 & 15.02  & 14.67 \\
      &   & w/ model ensemble & 9.00 & 10.76 & 13.78 & 12.02 \\
\cmidrule{2-7}      & \multirow{2}[2]{*}{ShallowCNN} & w/o model ensemble & 20.56 & 13.45 & 26.39  & 26.52 \\
      &   & w/ model ensemble & 19.78 & 13.36 & 24.06  & 20.93 \\
    \bottomrule
    \end{tabular}
\end{table*}

\begin{table*}[htbp] \footnotesize   \centering \setlength{\tabcolsep}{.8mm}
\tabcolsep 3pt
  \caption{Test BCAs of different UID models in ablation analysis on MI2.}\label{tab:ablaMI2}
    \begin{tabular}{c|c|c|c|c|c|c}
    \toprule
     \multirow{2}{*}{Approach} & \multirow{2}{*}{Model} & \multirow{2}{*}{Setting} & Adversarial & Surface & Temporal  & Temporal \\
&  & & Training & Laplacian & Shift  & Recombination \\
    \midrule
    \multirow{6}[6]{*}{SN} & \multirow{2}[2]{*}{EEGNet} & w/o channel variation & 3.18 & 19.43 & 6.06 & 32.45 \\
      &   & w/ channel variation & 3.31 & 1.92 & 2.78  & 7.01 \\
\cmidrule{2-7}      & \multirow{2}[2]{*}{DeepCNN} & w/o channel variation & 2.98 & 33.51 & 37.72 &  60.89 \\
      &   & w/ channel variation & 3.30 & 2.34 & 14.16  & 15.44 \\
\cmidrule{2-7}      & \multirow{2}[2]{*}{ShallowCNN} & w/o channel variation & 6.40 & 58.99 & 52.03 & 67.95 \\
      &   & w/ channel variation & 9.52 & 2.51 & 8.99  & 9.61 \\
    \midrule
    \multirow{6}[6]{*}{EMIN} & \multirow{2}[2]{*}{EEGNet} & w/o $\mathrm{trans(\cdot)}$ & 6.13 & 0.89 & 23.67  & 21.55 \\
      &   & w/ $\mathrm{trans(\cdot)}$ & 7.97 & 0.98 & 2.91 & 3.93 \\
\cmidrule{2-7}      & \multirow{2}[2]{*}{DeepCNN} & w/o $\mathrm{trans(\cdot)}$ & 4.72 & 1.31 & 51.11 & 52.36 \\
      &   & w/ $\mathrm{trans(\cdot)}$ & 7.34 & 1.35 & 16.62 & 11.46 \\
\cmidrule{2-7}      & \multirow{2}[2]{*}{ShallowCNN} & w/o $\mathrm{trans(\cdot)}$ & 7.91 & 1.03 & 51.44 &  51.06 \\
      &   & w/ $\mathrm{trans(\cdot)}$ & 11.89 & 1.11 & 16.15 &  15.71 \\
    \midrule
    \multirow{6}[6]{*}{EMAX} & \multirow{2}[2]{*}{EEGNet} & w/o model ensemble & 5.89 & 1.23 & 3.86 & 4.11 \\
      &   & w/ model ensemble & 5.15 & 1.04 & 2.14 & 2.20 \\
\cmidrule{2-7}      & \multirow{2}[2]{*}{DeepCNN} & w/o model ensemble & 5.86 & 1.38 & 17.35 &  18.51 \\
      &   & w/ model ensemble & 5.78 & 1.39 & 8.11 & 8.29 \\
\cmidrule{2-7}      & \multirow{2}[2]{*}{ShallowCNN} & w/o model ensemble & 23.72 & 2.81 & 26.24 & 31.64 \\
      &   & w/ model ensemble & 19.26 & 2.00 & 13.69 & 14.14 \\
    \bottomrule
    \end{tabular}
\end{table*}

\begin{table*}[htbp]\footnotesize  \centering \setlength{\tabcolsep}{.8mm}
\tabcolsep 3pt
  \caption{Test BCAs of different UID models in ablation analysis on P300.}\label{tab:ablaP300}
    \begin{tabular}{c|c|c|c|c|c|c}
    \toprule
     \multirow{2}{*}{Approach} & \multirow{2}{*}{Model} & \multirow{2}{*}{Setting} & Adversarial & Surface & Temporal  & Temporal \\
&  & & Training & Laplacian & Shift  & Recombination \\
    \midrule
    \multirow{6}[6]{*}{SN} & \multirow{2}[2]{*}{EEGNet} & w/o channel variation & 12.16 & 30.57 & 14.85 & 55.81 \\
      &   & w/ channel variation & 13.55 & 12.93 & 13.88  & 11.91 \\
\cmidrule{2-7}      & \multirow{2}[2]{*}{DeepCNN} & w/o channel variation & 11.27 & 30.67 & 15.23 &  81.97 \\
      &   & w/ channel variation & 10.94 & 10.78 & 16.89 & 18.73 \\
\cmidrule{2-7}      & \multirow{2}[2]{*}{ShallowCNN} & w/o channel variation & 13.72 & 30.51 & 22.60  & 65.92 \\
      &   & w/ channel variation & 22.57 & 9.87 & 19.20  & 20.04 \\
    \midrule
    \multirow{6}[6]{*}{EMIN} & \multirow{2}[2]{*}{EEGNet} & w/o $\mathrm{trans(\cdot)}$ & 11.03 & 10.37 & 14.32 &  10.52 \\
      &   & w/ $\mathrm{trans(\cdot)}$ & 9.43 & 9.31 & 11.81 &  8.61 \\
\cmidrule{2-7}      & \multirow{2}[2]{*}{DeepCNN} & w/o $\mathrm{trans(\cdot)}$ & 10.89 & 10.13 & 12.75 & 15.02 \\
      &   & w/ $\mathrm{trans(\cdot)}$ & 10.44 & 9.38 & 12.27 & 11.69 \\
\cmidrule{2-7}      & \multirow{2}[2]{*}{ShallowCNN} & w/o $\mathrm{trans(\cdot)}$ & 10.49 & 10.07 & 17.86 & 13.81 \\
      &   & w/ $\mathrm{trans(\cdot)}$ & 13.54 & 10.28 & 13.03 & 10.91 \\
    \midrule
    \multirow{6}[6]{*}{EMAX} & \multirow{2}[2]{*}{EEGNet} & w/o model ensemble & 7.92 & 8.59 & 9.28 & 15.94 \\
      &   & w/ model ensemble & 7.73 & 9.30 & 10.42 & 9.39 \\
\cmidrule{2-7}      & \multirow{2}[2]{*}{DeepCNN} & w/o model ensemble & 6.38 & 9.36 & 11.42 & 12.06 \\
      &   & w/ model ensemble & 8.35 & 10.05 & 12.52  & 8.72 \\
\cmidrule{2-7}      & \multirow{2}[2]{*}{ShallowCNN} & w/o model ensemble & 11.92 & 9.40 & 13.60 & 14.34 \\
      &   & w/ model ensemble & 11.31 & 9.91 & 13.09 &  14.10 \\
    \bottomrule
    \end{tabular}
\end{table*}

\subsection{Parameter Sensibility Analysis} \label{sect:psa}

\begin{figure*}\centering
\subfigure[]{\includegraphics[width=.99\linewidth,clip]{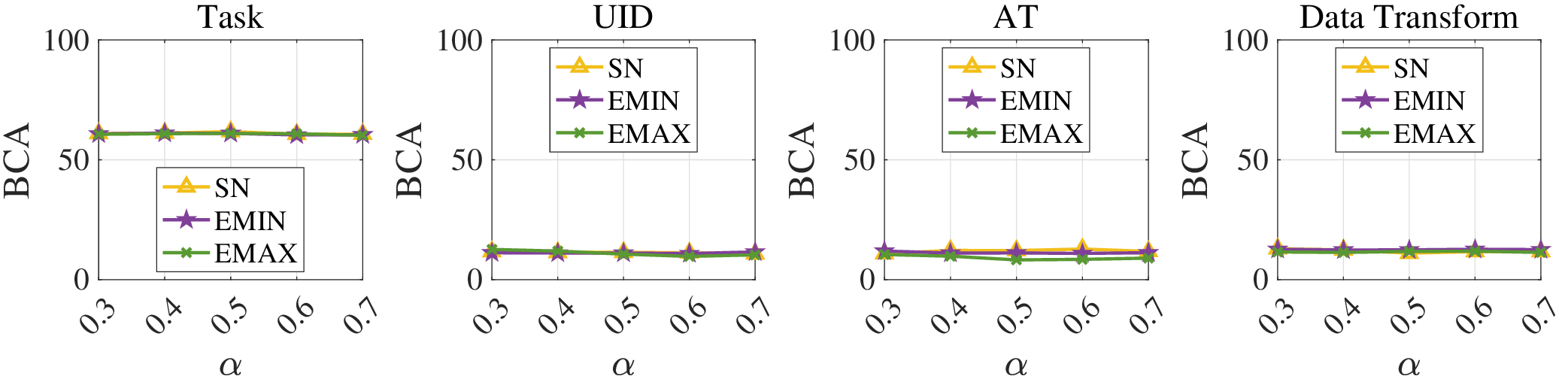}\label{fig:oa}}
\subfigure[]{\includegraphics[width=.99\linewidth,clip]{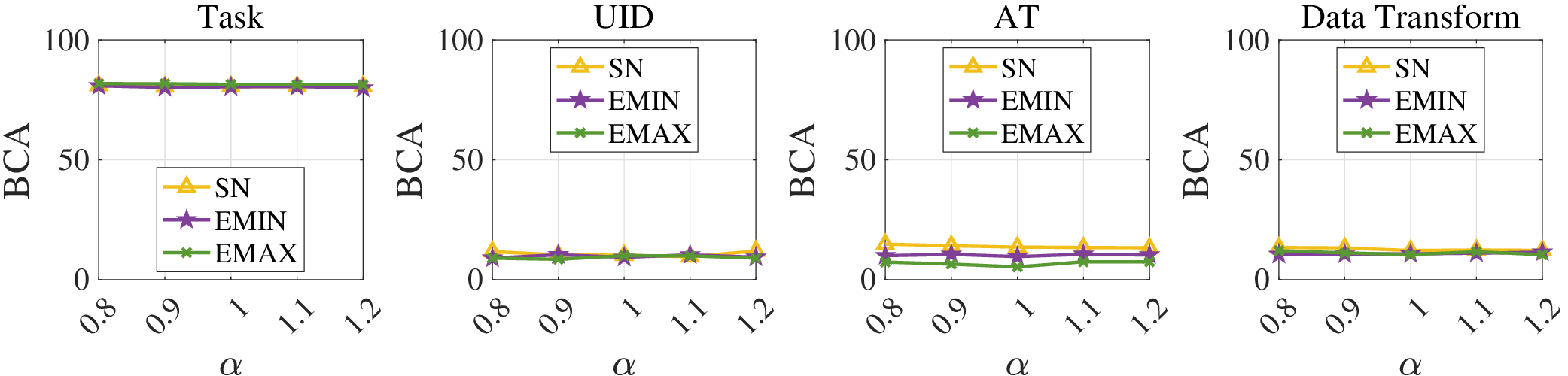}\label{fig:oa}}
\caption{Parameter sensitivity analysis on (a) MI1 and (b) P300.}\label{fig:sen}
\end{figure*}

We conducted experiments on the MI1 and P300 datasets and EEGNet to analyze the influence of perturbation amplitude $\alpha$ on the proposed three kinds of privacy-preserving perturbations. Models' BCAs on task, UID, under AT, and data transformation are shown in Figure~\ref{fig:sen}. Under a variety of amplitude values, three kinds of privacy-preserving perturbations we propose maintained good effectiveness and robustness (low UID BCA) without significantly affecting task classification performance.

When a broader range of perturbation amplitudes is considered, an intuitive conclusion can be drawn: smaller perturbation amplitudes make the privacy-preserving perturbations less noticeable, resulting in less impact on the EEG data. However, the perturbation amplitude should not be too small, as this could reduce robustness and render privacy protection ineffective. Conversely, if the amplitude is too large, it may interfere with the recognition of task-related information in the EEG data. Figure~6 shows that, within a certain range, privacy-preserving perturbations maintain a stable effect with no significant impact on the primary EEG classification performance while ensuring robustness against adversarial attacks and data transformations.

The appropriate range of the privacy-preserving perturbation amplitude may depend on factors such as the BCI paradigm, the type of privacy-preserving perturbation, and the machine learning models used. In practice, experiments can be conducted to assess the effects of different amplitudes and types of perturbations under various conditions before dataset release, allowing for the selection of the most suitable perturbation. An ideal privacy-preserving perturbation should maintain robustness across different conditions without significantly impacting the primary BCI classification performance.

\subsection{Comparison with our previous work} \label{sect:c}

Our previous work by Meng \emph{et al.} \cite{Meng2023} introduced perturbations to both the training and test data to prevent deep learning models from identifying private information in EEG data. This subsection demonstrates that it cannot prevent machine learning models from learning private information during the training phase, as mentioned in Introduction.

Table~\ref{tab:meng} presents the performance of the user-wise perturbation method in \cite{Meng2023} on three datasets and three networks when the test data was not perturbed. The hyperparameters were set according to the original paper \cite{Meng2023}. A comparison between the original UID BCAs and those obtained using the approach in \cite{Meng2023} shows that \cite{Meng2023} only provided good privacy protection for EEGNet on MI1, reducing the UID BCA from 87.93\% to 18.54\%.

In contrast, experimental results from previous sections demonstrated that the privacy-preserving perturbations proposed in this paper provided effective privacy protection across multiple cases, even without perturbation on the test data. They remained robust under AT, data transformations, and across both deep and traditional machine learning models, indicating that our proposed approach is more effective and practical.

\begin{table}[htbp]
  \centering \setlength{\tabcolsep}{1.4mm}
  \small
  \caption{Test BCAs of BCI task classifiers and UID models trained on the original (unperturbed) EEG data and those with user-wise perturbations proposed in \cite{Meng2023}.}
    \begin{tabular}{c|c|c|cc|cc}
    \toprule
    \multirow{2}[2]{*}{Dataset} & \multirow{2}[2]{*}{\# users} & \multirow{2}[2]{*}{Model} & \multicolumn{2}{c|}{Original} & \multicolumn{2}{c}{\cite{Meng2023}} \\
      &   &   & Task & UID & Task & UID \\
    \midrule
    \multirow{3}[2]{*}{MI1} & \multirow{3}[2]{*}{9} & EEGNet & 61.85 & 87.93 & 61.10 & 18.54 \\
      &   & DeepCNN & 63.82 & 92.79 & 62.99 & 85.43 \\
      &   & ShallowCNN & 66.10 & 99.02 & 66.81 & 71.77 \\
    \midrule
    \multirow{3}[2]{*}{MI2} & \multirow{3}[2]{*}{109} & EEGNet & 71.07 & 53.99 & 70.42 & 53.14 \\
      &   & DeepCNN & 72.76 & 72.46 & 73.06 & 71.59 \\
      &   & ShallowCNN & 72.71 & 87.39 & 72.17 & 87.68 \\
    \midrule
    \multirow{3}[2]{*}{ERN} & \multirow{3}[2]{*}{16} & EEGNet & 65.26 & 41.43 & 64.96 & 46.98 \\
      &   & DeepCNN & 64.11 & 65.96 & 64.13 & 65.45 \\
      &   & ShallowCNN & 65.19 & 45.44 & 65.39 & 45.83 \\
    \bottomrule
    \end{tabular}%
  \label{tab:meng}%
\end{table}%

\section{Conclusions} \label{sect:CFR}

This paper has proposed four types of user-wise privacy-preserving perturbations to construct user identity unlearnable EEG training data, each with unique characteristics:
\begin{enumerate}
\item RAND is simple, almost parameter-free, and effective when the perturbed EEG training trails are used directly; however, its performance may be unstable under AT or data augmentation.
\item SN is training-free, but it requires manually designed coding. Moreover, its performance is slightly worse than EMIN and EMAX.
\item Both EMIN and EMAX require training, but they generally have the best performance, and are robust under AT and data preprocessing/transformation.
\end{enumerate}

This research protects user identity information in EEG, while maintaining the primary BCI task classification performance. Our future research will protect user identity information in EEG, while improving the primary BCI task classification performance.

\section{Reference}

\end{document}